\documentclass[aps,prl,twocolumn,groupedaddress,showpacs,amsmath,amssymb,floatfix,superscriptaddress]{revtex4}
\usepackage{graphicx}
\usepackage{times}
\usepackage{multirow}

\newcommand{\nuebar}{$\overline{\nu}_{e}$}
\newcommand{\DeltaMSq}{$\Delta m^{2}_{21}$}
\newcommand{\SolTheta}{$\theta_{12}$}
\newcommand{\ThetaParam}{$\tan^2 \theta_{12}$}
\newcommand{\LiHe}{$^9$Li/$^8$He}
\newcommand{\BN}{$^{12}$B/$^{12}$N}
\newcommand{\alphan}{$^{13}$C($\alpha$,n)$^{16}$O}

\newcommand{\Ugeo}{$^{238}$U}
\newcommand{\Thgeo}{$^{232}$Th}

\begin{document}

\title{ Precision Measurement of Neutrino Oscillation Parameters with KamLAND}

% All university affiliations addresses go here:
\newcommand{\tohoku}{\affiliation{Research Center for Neutrino
    Science, Tohoku University, Sendai 980-8578, Japan}}
\newcommand{\alabama}{\affiliation{Department of Physics and
    Astronomy, University of Alabama, Tuscaloosa, Alabama 35487, USA}}
\newcommand{\lbl}{\affiliation{Physics Department, University of
    California, Berkeley and \\ Lawrence Berkeley National Laboratory, 
Berkeley, California 94720, USA}}
\newcommand{\caltech}{\affiliation{W.~K.~Kellogg Radiation Laboratory,
    California Institute of Technology, Pasadena, California 91125, USA}}
\newcommand{\colostate}{\affiliation{Department of Physics, Colorado
    State University, Fort Collins, Colorado 80523, USA}}
\newcommand{\drexel}{\affiliation{Physics Department, Drexel
    University, Philadelphia, Pennsylvania 19104, USA}}
\newcommand{\hawaii}{\affiliation{Department of Physics and Astronomy,
    University of Hawaii at Manoa, Honolulu, Hawaii 96822, USA}}
\newcommand{\kansas}{\affiliation{Department of Physics,
    Kansas State University, Manhattan, Kansas 66506, USA}}
\newcommand{\lsu}{\affiliation{Department of Physics and Astronomy,
    Louisiana State University, Baton Rouge, Louisiana 70803, USA}}
\newcommand{\stanford}{\affiliation{Physics Department, Stanford
    University, Stanford, California 94305, USA}}
\newcommand{\ut}{\affiliation{Department of Physics and
    Astronomy, University of Tennessee, Knoxville, Tennessee 37996, USA}}
\newcommand{\tunl}{\affiliation{Triangle Universities Nuclear
    Laboratory, Durham, North Carolina 27708, USA and \\
Physics Departments at Duke University, North Carolina Central University,
and the University of North Carolina at Chapel Hill}}
\newcommand{\wisc}{\affiliation{Department of Physics, University
    of Wisconsin, Madison, Wisconsin 53706, USA}}  
\newcommand{\cnrs}{\affiliation{CEN Bordeaux-Gradignan, IN2P3-CNRS and
    University Bordeaux I, F-33175 Gradignan Cedex, France}}

% Put Present addresses here:
\newcommand{\aticrrnow}{\altaffiliation{Present address: ICRR, 
    University of Tokyo, Gifu, Japan}}
\newcommand{\aticeppnow}{\altaffiliation{Present address: ICEPP,
    University of Tokyo, Tokyo, Japan}}
\newcommand{\atimperialnow}{\altaffiliation{Present address: Imperial
    College London, UK}}
\newcommand{\atlanlnow}{\altaffiliation{Present address: LANL, Los
    Alamos, NM 87545, USA}}
\newcommand{\atiasnow}{\altaffiliation{Present address: School of
    Natural Sciences, Institute for Advanced Study, Princeton, NJ
    08540, USA}}
\newcommand{\atksunow}{\altaffiliation{Present address: KSU, Manhattan, KS 66506, USA}}
\newcommand{\atdubnanow}{\altaffiliation{Present address: DLNP, JINR,
Dubna, Russia}}
\newcommand{\atokayamanow}{\altaffiliation{Present address: 
Center of Quantum Universe, Okayama University, Okayama 700-8530,
Japan}}

\newcommand{\atregisnow}{\altaffiliation{Present address: Regis University, Denver, CO 80221, USA}}
\newcommand{\atfnalnow}{\altaffiliation{Present address: FNAL, Batavia, IL 60510, USA}}
\newcommand{\atsnolabnow}{\altaffiliation{Present address: SNOLAB,
Lively, ON P3Y 1M3, Canada}}
\newcommand{\atllnlnow}{\altaffiliation{Present address: LLNL, Livermore, CA 94550, USA}}

% Tohoku
\author{S.~Abe}\tohoku
\author{T.~Ebihara}\tohoku
\author{S.~Enomoto}\tohoku
\author{K.~Furuno}\tohoku
\author{Y.~Gando}\tohoku
\author{K.~Ichimura}\tohoku
\author{H.~Ikeda}\tohoku
\author{K.~Inoue}\tohoku
\author{Y.~Kibe}\tohoku
\author{Y.~Kishimoto}\tohoku
\author{M.~Koga}\tohoku
\author{A.~Kozlov}\tohoku
\author{Y.~Minekawa}\tohoku
\author{T.~Mitsui}\tohoku
\author{K.~Nakajima}\atokayamanow\tohoku % Kyo Nakajima
\author{K.~Nakajima}\tohoku
\author{K.~Nakamura}\tohoku
\author{M.~Nakamura}\tohoku
\author{K.~Owada}\tohoku
\author{I.~Shimizu}\tohoku
\author{Y.~Shimizu}\tohoku
\author{J.~Shirai}\tohoku
\author{F.~Suekane}\tohoku
\author{A.~Suzuki}\tohoku
\author{Y.~Takemoto}\tohoku
\author{K.~Tamae}\tohoku
\author{A.~Terashima}\tohoku
\author{H.~Watanabe}\tohoku
\author{E.~Yonezawa}\tohoku
\author{S.~Yoshida}\tohoku
%
% Alabama
\author{J.~Busenitz}\alabama
\author{T.~Classen}\alabama
\author{C.~Grant}\alabama
\author{G.~Keefer}\alabama
\author{D.S.~Leonard}\alabama
\author{D.~McKee}\alabama
\author{A.~Piepke}\alabama
%
% LBL and UC Berkeley
\author{M.P.~Decowski}\lbl
\author{J.A.~Detwiler}\lbl
\author{S.J.~Freedman}\lbl
\author{B.K.~Fujikawa}\lbl
\author{F.~Gray}\atregisnow\lbl
\author{E.~Guardincerri}\lbl
\author{L.~Hsu}\atfnalnow\lbl
\author{R.~Kadel}\lbl
\author{C.~Lendvai}\lbl
\author{K.-B.~Luk}\lbl
\author{H.~Murayama}\lbl
\author{T.~O'Donnell}\lbl
\author{H.M.~Steiner}\lbl
\author{L.A.~Winslow}\lbl
%
% Caltech
\author{D.A.~Dwyer}\caltech
\author{C.~Jillings}\atsnolabnow\caltech
\author{C.~Mauger}\caltech
\author{R.D.~McKeown}\caltech
\author{P.~Vogel}\caltech
\author{C.~Zhang}\caltech
%
% Colorado State
\author{B.E.~Berger}\colostate
%
% Drexel
\author{C.E.~Lane}\drexel
\author{J.~Maricic}\drexel
\author{T.~Miletic}\drexel
%
% Hawaii
\author{M.~Batygov}\hawaii
\author{J.G.~Learned}\hawaii
\author{S.~Matsuno}\hawaii
\author{S.~Pakvasa}\hawaii
%
% KSU
\author{J.~Foster}\kansas
\author{G.A.~Horton-Smith}\kansas
\author{A.~Tang}\kansas
%
% LSU
\author{S.~Dazeley}\atllnlnow\lsu
%
% Stanford
\author{K.E.~Downum}\stanford
\author{G.~Gratta}\stanford
\author{K.~Tolich}\stanford
%
% UT
\author{W.~Bugg}\ut
\author{Y.~Efremenko}\ut
\author{Y.~Kamyshkov}\ut
\author{O.~Perevozchikov}\ut
%
% TUNL
\author{H.J.~Karwowski}\tunl
\author{D.M.~Markoff}\tunl
\author{W.~Tornow}\tunl
%
% Wisconscin
\author{K.M.~Heeger}\wisc
% 
% CNRS
\author{F.~Piquemal}\cnrs
\author{J.-S.~Ricol}\cnrs

\collaboration{The KamLAND Collaboration}\noaffiliation

\date{\today%\\
%\textbf{version 4.9}
}

\begin{abstract}
  The KamLAND experiment has determined a precise value for the
  neutrino oscillation parameter \DeltaMSq\ and stringent constraints
  on \SolTheta. The exposure to nuclear reactor anti-neutrinos is increased
  almost fourfold over previous results to 2.44$\times$10$^{32}$\,proton-yr
  due to longer livetime and an enlarged fiducial volume. An
  undistorted reactor \nuebar\ energy spectrum is now rejected at
  $>$5$\sigma$. Analysis of the reactor spectrum above the inverse beta decay
energy treshold, and including geo-neutrinos, gives a best-fit at \DeltaMSq\,=\,$7.58^{+0.14}_{-0.13}(\text{stat})^{+0.15}_{-0.15}(\text{syst})\times10^{-5}$\,eV$^{2}$ and
  \ThetaParam\,=\,$0.56^{+0.10}_{-0.07}(\text{stat})^{+0.10}_{-0.06}(\text{syst})$\@. Local $\Delta \chi^2$-minima at
  higher and lower \DeltaMSq\ are disfavored at $>$4$\sigma$. Combining
  with solar neutrino data, we obtain
  \DeltaMSq\,=\,$7.59^{+0.21}_{-0.21}\times10^{-5}$\,eV$^{2}$ and
  \ThetaParam\,=\,$0.47^{+0.06}_{-0.05}$\@.
\end{abstract}


\maketitle
Experiments studying atmospheric, solar, reactor and accelerator
neutrinos provide compelling evidence for neutrino mass and
oscillation. The Kamioka Liquid scintillator Anti-Neutrino Detector
(KamLAND) investigates neutrino oscillation parameters by observing
electron anti-neutrinos (\nuebar) emitted from distant nuclear
reactors. Previously, KamLAND announced the first evidence of \nuebar\
disappearance~\cite{1stResult}, followed by direct evidence for
neutrino oscillation by observing distortion of the reactor \nuebar\
energy spectrum~\cite{2ndResult}.  More recently, KamLAND showed the
first indication of geologically produced anti-neutrinos
(geo-neutrinos) from radioactive decay in the Earth~\cite{GeoNature},
possibly a unique tool for geology.

This Letter presents a precise measurement of \DeltaMSq\ and new
constraints on \SolTheta\ based on data collected from
March 9, 2002 to May 12, 2007, including data used earlier~\cite{1stResult,2ndResult}. We have enlarged the fiducial
volume radius from 5.5\,m to 6\,m and collected significantly more
data; the total exposure is 2.44$\times$10$^{32}$\,proton-yr
(2881\,ton-yr).  We have expanded the analysis to the full
reactor \nuebar\ energy spectrum and reduced the systematic
uncertainties in the number of target protons and the background. We
now observe almost two complete oscillation cycles in the \nuebar\
spectrum and extract more precise values of the oscillation
parameters. 

KamLAND is at the site of the former Kamiokande experiment at a depth
of $\sim$2700\,m water equivalent. The heart of the detector is
1\,kton of highly purified liquid scintillator (LS) enclosed in an
EVOH/nylon balloon suspended in purified mineral oil. The LS consists
of 80\% dodecane, 20\% pseudocumene and 1.36\,$\pm$\,0.03\,g/l of
PPO~\cite{PreviouslyAFootnote}. The anti-neutrino detector is inside
an 18-m-diameter stainless steel sphere. An array of 1879
50-cm-diameter photomultiplier tubes (PMTs) is mounted on the inner
surface of the sphere. 554 of these are reused from the Kamiokande
experiment, while the remaining 1325 are a faster version masked to 17
inches. A 3.2-kton cylindrical water-Cherenkov outer detector (OD),
surrounding the containment sphere, provides shielding and operates as
an active cosmic-ray veto detector.

Electron anti-neutrinos are detected via inverse $\beta$-decay,
$\overline{\nu}_e + p\rightarrow e^+ + n$, with a 1.8\,MeV
threshold. The prompt scintillation light from the $e^+$ gives a
measure of the \nuebar\ energy,
$E_{\overline{\nu}_e}\,\simeq\,E_{\text{p}} + \overline{E}_n +
0.8$\,MeV, where $E_{\text{p}}$ is the prompt event energy including
the positron kinetic and annihilation energy, and $\overline{E}_n$ is
the average neutron recoil energy, {\cal O(10\,keV)}. The mean neutron
capture time is 207.5\,$\pm$\,2.8\,$\mu$s. More than 99\% capture on
free protons, producing a 2.2\,MeV $\gamma$ ray.

KamLAND is surrounded by 55 Japanese nuclear power reactor units, each
an isotropic \nuebar\ source. The reactor operation records, including
thermal power generation, fuel burnup, and exchange and enrichment
logs, are provided by a consortium of Japanese electric power
companies. This information, combined with publicly available world
reactor data, is used to calculate the instantaneous fission rates
using a reactor model~\cite{ReactorModel}. Only four isotopes
contribute significantly to the \nuebar\ spectra; the ratios of the
fission yields averaged over the entire data taking period are:
$^{235}$U\,:\,$^{238}$U\,:\,$^{239}$Pu\,:\,$^{241}$Pu\,=\,0.570:\,0.078:\,0.295:\,0.057. The
emitted \nuebar\ energy spectrum is calculated using the \nuebar\
spectra inferred from Ref.~\cite{Spectra}, while the spectral
uncertainty is evaluated from Ref.~\cite{SpectraUncert}. We also
include contributions from the long-lived fission daughters $^{90}$Sr,
$^{106}$Ru, and $^{144}$Ce~\cite{Kop01}.

We recently commissioned an ``off-axis'' calibration system capable of
positioning radioactive sources away from the central vertical axis of
the detector. The measurements indicate that the vertex
reconstruction systematic deviations are radius- and
zenith-angle-dependent, but smaller
than 3\,cm and independent of azimuthal angle.  The
fiducial volume (FV) is known to 1.6\% uncertainty up to 5.5\,m
using the off-axis calibration system. The position distribution of
the $\beta$-decays of muon-induced \BN\ confirms this
with 4.0\% uncertainty by comparing the number of events
inside 5.5\,m to the number produced in the full LS volume. The \BN\
event ratio is used to establish the uncertainty between 5.5\,m and
6\,m, resulting in a combined 6-m-radius FV uncertainty of
1.8\%.

\begin{table}[tb]
  \caption{Estimated systematic uncertainties relevant for the neutrino
    oscillation parameters \DeltaMSq\ and \SolTheta. }
\label{tab:sys}
\begin{tabular*}{8.5cm}{@{\extracolsep{\fill}}c|llll}
\hline\hline
 & Detector-related (\%) &  & Reactor-related (\%) &  \\ \hline
\DeltaMSq & Energy scale & 1.9 & \nuebar-spectra~\cite{SpectraUncert} & 0.6 \\ \hline
\multirow{4}{*}{Event rate} & Fiducial volume & 1.8 & \nuebar-spectra & 2.4 \\
 & Energy threshold & 1.5 & Reactor power & 2.1 \\
 & Efficiency & 0.6 & Fuel composition & 1.0 \\
 & Cross section & 0.2 & Long-lived nuclei & 0.3 \\
\hline\hline
\end{tabular*}
\end{table}

Off-axis calibration measurements and numerous central-axis
deployments of $^{60}$Co, $^{68}$Ge, $^{203}$Hg, $^{65}$Zn,
$^{241}$Am$^{9}$Be, $^{137}$Cs and $^{210}$Po$^{13}$C radioactive
sources established the event reconstruction performance. The vertex
reconstruction resolution is $\sim$12\,cm/$\sqrt{E(\text{MeV})}$ and
the energy resolution is $6.5\%/\sqrt{E(\text{MeV})}$. The
scintillator response is corrected for the non-linear effects from
quenching and Cherenkov light production. The systematic variation of
the energy reconstruction over the data-set give an absolute
energy-scale uncertainty of 1.4\%; the distortion of the energy scale
results in a 1.9\% uncertainty on \DeltaMSq, while the uncertainty at
the analysis threshold gives a 1.5\% uncertainty on the event
rate. Table~\ref{tab:sys} summarizes the systematic uncertainties. The
total uncertainty on \DeltaMSq\ is 2.0\%, while the uncertainty on the
expected event rate, which primarily affects \SolTheta, is 4.1\%.

For the analysis we require
0.9\,MeV$<E_{\text{p}}<$\,8.5\,MeV\@. 
The delayed energy,
$E_{\text{d}}$, must satisfy 1.8\,MeV\,$<\,E_{\text{d}}\,<$\,2.6\,MeV
or 4.0\,MeV\,$<\,E_{\text{d}}\,<$\,5.8\,MeV, corresponding to the
neutron-capture $\gamma$ energies for p and $^{12}$C,
respectively.  The time difference ($\Delta T$) and distance ($\Delta R$) between the prompt event and delayed neutron capture are selected
to be 
0.5\,$\mu$s$<\Delta T<$\,1000\,$\mu$s and 
$\Delta R$\,$<$\,2\,m. The prompt and delayed radial distance from the
detector center ($R_{\text{p}}$, $R_{\text{d}}$) must be $<$6\,m. 

\begin{figure}
\includegraphics[width=6cm,angle=270]{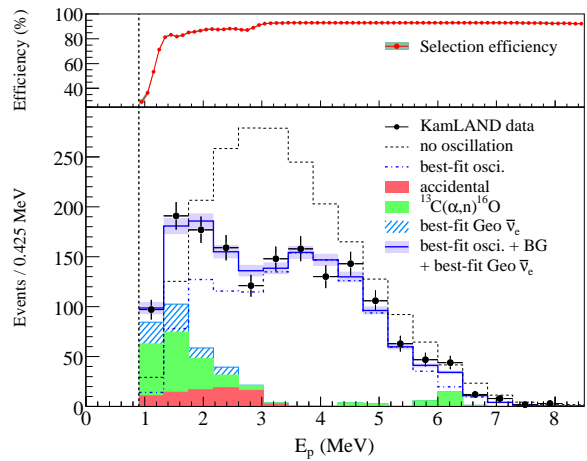}
\caption[] {Prompt event energy spectrum of \nuebar\ candidate
  events. All histograms corresponding to reactor spectra and expected
  backgrounds incorporate the energy-dependent selection efficiency
  (top panel). The shaded background and geo-neutrino histograms are
  cumulative. Statistical uncertainties are shown for the data; the band on
  the blue histogram indicates the event rate systematic uncertainty.}
\label{fig:E-spectrum}
\end{figure}

Accidental coincidences increase near the balloon surface
($R$\,=\,6.5\,m), reducing the signal-to-background ratio.  We use
constraints on event characteristics to suppress accidental
backgrounds while maintaining high efficiency. We construct a
probability density function (PDF) for accidental coincidence events,
$f_{acc}(E_{\text{p}},E_{\text{d}},\Delta R,\Delta T,
R_{\text{p}},R_{\text{d}})$, by pairing events in a 10-ms-to-20-s
delayed-coincidence window. A PDF for the \nuebar\ signal,
$f_{\overline{\nu}_{e}}(E_{\text{p}},E_{\text{d}},\Delta R,\Delta T,
R_{\text{p}},R_{\text{d}})$, is constructed from a Monte Carlo
simulation of the prompt and delayed events using the measured neutron
capture time and detector response. For the $E_{\text{p}}$
distribution in $f_{\overline{\nu}_{e}}$, we choose an
oscillation-free reactor spectrum including a contribution from
geo-neutrinos estimated from Ref.~\cite{GeoReference}. A discriminator
value,
$L=\frac{f_{\overline{\nu}_{e}}}{f_{\overline{\nu}_{e}}+f_{acc}}$, is
calculated for each candidate pair that passes the earlier cuts. We
establish a selection value $L_{i}^{cut}$ in $E_{\text{p}}$ bins of
0.1\,MeV, where $L_{i}^{cut}$ is the value of $L$ at which the
figure-of-merit, $\frac{S_i}{\sqrt{S_i+B_i}}$ is maximal. $S_i$ is the
number of Monte Carlo signal events in the {\it i}th energy bin with
$L\,>\,L_{i}^{cut}$. $B_i$ is calculated similarly using the
accidental coincidence event pairs. The choice of the $E_{\text{p}}$
distribution in $f_{\overline{\nu}_{e}}$ affects only the
discrimination power of the procedure; substituting the
oscillation-free reactor spectrum by an oscillated spectrum with the
parameters from Ref.~\cite{2ndResult} changes our oscillation
parameter results by less than 0.2$\sigma$. The selection efficiency
$\epsilon(E_{\text{p}})$ is estimated from the fraction of selected
coincidence events relative to the total generated in R\,$<$\,6\,m in
the simulation, see Fig.~\ref{fig:E-spectrum}(top).

\begin{table}[tb]
\caption{Estimated backgrounds after selection efficiencies.}
\label{tab:bg}
\begin{tabular}{lll}
\hline\hline
Background & &Contribution \\ \hline
Accidentals & & 80.5\,$\pm$\,0.1 \\
\LiHe\ & & 13.6\,$\pm$\,1.0 \\
\multicolumn{2}{l}{Fast neutron \& Atmospheric $\nu$} & $<$9.0 \\
$^{13}$C($\alpha$,n)$^{16}$O$_{gs}$, & np $\rightarrow$ np & 157.2\,$\pm$\,17.3 \\
$^{13}$C($\alpha$,n)$^{16}$O$_{gs}$, & $^{12}$C(n,n$'$)$^{12}$C$^{*}$ (4.4\,MeV $\gamma$) & 6.1\,$\pm$\,0.7 \\
\multicolumn{2}{l}{\alphan\ 1$^{\text{st}}$ exc. state (6.05\,MeV e$^{+}$e$^{-}$)} & 15.2\,$\pm$\,3.5 \\
\multicolumn{2}{l}{\alphan\ 2$^{\text{nd}}$ exc. state (6.13\,MeV $\gamma$)} & 3.5\,$\pm$\,0.2 \\
\hline
Total & & 276.1\,$\pm$\,23.5 \\ 
\hline\hline
\end{tabular}
\end{table}

The dominant background is caused by \alphan\ reactions from
$\alpha$-decay of $^{210}$Po, a daughter of $^{222}$Rn introduced into
the LS during construction. We estimate that there are
(5.56\,$\pm$\,0.22)\,$\times\,10^{9}$ $^{210}$Po $\alpha$-decays. The
\alphan\ reaction results in neutrons with energies up to 7.3\,MeV,
but most of the scintillation energy spectrum is quenched below
2.7\,MeV. In addition, $^{12}$C(n,n$'$)$^{12}$C$^{*}$, and the
1$^{\text{st}}$ and 2$^{\text{nd}}$ excited states of $^{16}$O produce
signals in coincidence with the scattered neutron but the cross
sections are not known precisely. A $^{210}$Po$^{13}$C source was
employed to study the \alphan\ reaction and tune a simulation using
the cross sections from Ref.~\cite{Har05,JENDL}. We find that the
cross sections for the excited $^{16}$O states from Ref.~\cite{JENDL}
agree with the $^{210}$Po$^{13}$C data after scaling the
1$^{\text{st}}$ excited state by 0.6; the 2$^{\text{nd}}$ excited
state requires no scaling. For the ground-state we use the cross
section from Ref.~\cite{Har05} and scale by 1.05. Including the
$^{210}$Po decay-rate, we assign an uncertainty of 11\% for the
ground-state and 20\% for the excited states. Accounting for
$\epsilon(E_{\text{p}})$, there should be 182.0\,$\pm$\,21.7 \alphan\
events in the data.

To mitigate background arising from the cosmogenic beta
delayed-neutron emitters $^9$Li and $^8$He, we apply a 2\,s veto
within a 3-m-radius cylinder around well-identified muon tracks
passing through the LS. For muons that either deposit a large amount
of energy or cannot be tracked, we apply a 2\,s veto of the full
detector. We estimate that 13.6\,$\pm$\,1.0 events from \LiHe\ decays
remain by fitting the time distribution of identified \LiHe\ since the
prior muons. Spallation-produced neutrons are suppressed with a 2\,ms
full-volume veto after a detected muon. Some neutrons are produced by
muons that are undetected by the OD or miss the OD but interact in the
nearby rock. These neutrons can scatter and capture in the LS,
mimicking the \nuebar\ signal. We also expect background events from
atmospheric neutrinos. The energy spectrum of these backgrounds is
assumed to be flat to at least 30\,MeV based on a simulation
following~\cite{G4SpallValid}. The atmospheric $\nu$
spectrum~\cite{AtmoSpec} and interactions were modeled using
NUANCE~\cite{NUANCE}. We expect fewer than 9 neutron and atmospheric
$\nu$ events in the data-set. We observe 15 events in the energy range
8.5 -- 30\,MeV, consistent with the limit reported
previously~\cite{HiNuebar}.

The accidental coincidence background above 0.9\,MeV is measured with
a 10-ms-to-20-s delayed-coincidence window to be 80.5\,$\pm$\,0.1
events. Other backgrounds from ($\gamma$,n) interactions and
spontaneous fission are negligible.

Anti-neutrinos produced in the decay chains of \Thgeo\ and \Ugeo\ in
the Earth's interior are limited to prompt energies below
2.6\,MeV. The expected geo-neutrino flux at the KamLAND location is
estimated with a geological reference model~\cite{GeoReference}, which assumes a
radiogenic heat production rate of 16\,TW from the U and Th-decay
chains. The calculated \nuebar\ fluxes for U and Th-decay, including a
suppression factor of 0.57 due to neutrino oscillation,
are 2.24$\times$10$^{6}$\,cm$^{-2}$s$^{-1}$ (56.6 events) and
1.90$\times$10$^{6}$\,cm$^{-2}$s$^{-1}$ (13.1 events),
respectively. 

With no \nuebar\ disappearance, we expect
2179\,$\pm$\,89\,(syst) events from reactors. The backgrounds in the reactor
energy region listed in Table~\ref{tab:bg} sum to 276.1\,$\pm$\,23.5; we
also expect geo-neutrinos. We observe 1609 events.

Figure~\ref{fig:E-spectrum} shows the prompt energy spectrum of
selected \nuebar\ events and the fitted backgrounds. The
unbinned data is assessed with a maximum likelihood fit to two-flavor
neutrino oscillation (with $\theta_{13}$\,=\,0), simultaneously
fitting the geo-neutrino contribution. The method incorporates the
absolute time of the event and accounts for time variations in the
reactor flux. Earth-matter oscillation effects are included. The
best-fit is shown in Fig.~\ref{fig:E-spectrum}. The joint confidence
intervals give \DeltaMSq\,=\,$7.58^{+0.14}_{-0.13}(\text{stat})^{+0.15}_{-0.15}(\text{syst})\times10^{-5}$\,eV$^{2}$ and
\ThetaParam\,=\,$0.56^{+0.10}_{-0.07}(\text{stat})^{+0.10}_{-0.06}(\text{syst})$ for \ThetaParam$<$1. A scaled reactor
spectrum with no distortion from neutrino oscillation is excluded at
more than 5$\sigma$. An independent analysis using cuts similar to
Ref.~\cite{2ndResult} gives
\DeltaMSq\,=\,$7.66^{+0.22}_{-0.20}\times10^{-5}$\,eV$^{2}$ and
\ThetaParam\,=\,$0.52^{+0.16}_{-0.10}$.

\begin{figure}
\includegraphics[width=7.75cm]{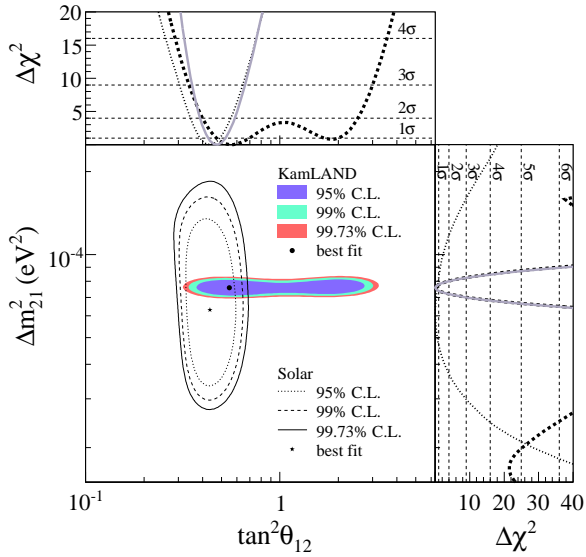}
\caption[] {Allowed region for neutrino oscillation parameters from
  KamLAND and solar neutrino experiments. The side-panels show the
  $\Delta \chi^{2}$-profiles for KamLAND (dashed) and
  solar experiments (dotted) individually, as well as the combination
  of the two (solid).}
\label{fig:osc-analysis}
\end{figure}

The allowed contours in the neutrino oscillation parameter space,
including $\Delta \chi^2$-profiles, are shown in
Fig.~\ref{fig:osc-analysis}. Only the so-called LMA-I region remains,
while other regions previously allowed by KamLAND at $\sim$2.2$\sigma$
are disfavored at more than 4$\sigma$. For three-neutrino
oscillation, the data give the same result for \DeltaMSq, but
a slightly larger uncertainty on \SolTheta. Incorporating the results of
SNO~\cite{snosalt} and solar flux experiments~\cite{solarflux} in a
two-neutrino analysis with KamLAND assuming CPT invariance, gives
\DeltaMSq\,=\,$7.59^{+0.21}_{-0.21}\times10^{-5}$\,eV$^{2}$ and
\ThetaParam\,=\,$0.47^{+0.06}_{-0.05}$.

To determine the number of geo-neutrinos, we fit the
normalization of the \nuebar\ energy spectrum from the U and Th-decay
chains simultaneously with the neutrino oscillation parameters using the KamLAND and solar data. There is a strong anti-correlation between the U and
  Th-decay chain geo-neutrinos and an unconstrained fit of the individual contributions does not give meaningful results. Fixing
the Th/U mass ratio to 3.9 from planetary data~\cite{ThURatio}, we
obtain a combined U+Th best-fit value of
(4.4\,$\pm$\,1.6)$\times$10$^{6}$\,cm$^{-2}$s$^{-1}$ (73\,$\pm$\,27
events), in agreement with the reference model.

The KamLAND data, together with the solar $\nu$ data, set an upper
limit of 6.2\,TW (90\% C.L.) for a \nuebar\ reactor source at the
Earth's center~\cite{GeoReactor}, assuming that the reactor produces a
spectrum identical to that of a slow neutron artificial reactor.

\begin{figure}
\includegraphics[width=5.5cm,angle=270]{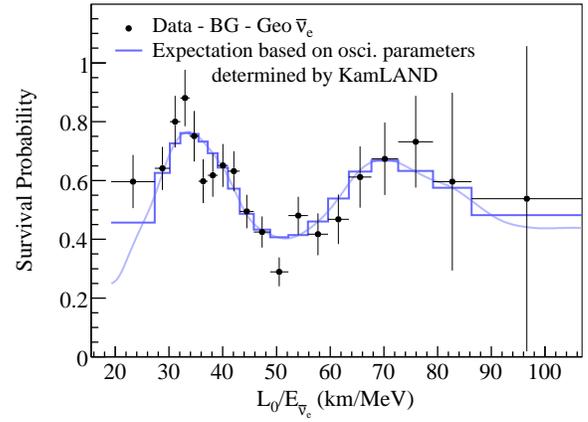}
\caption[] {Ratio of the background and geo-neutrino-subtracted
  \nuebar\ spectrum to the expectation for no-oscillation as a
  function of $L_{0}/E$. $L_{0}$ is the effective baseline taken as a
  flux-weighted average ($L_{0}$\,=\,180\,km). The energy bins are
  equal probability bins of the best-fit including all backgrounds
  (see Fig.~\ref{fig:E-spectrum}). The histogram and curve show the
  expectation accounting for the distances to the individual reactors,
  time-dependent flux variations and efficiencies. The error bars are
  statistical only and do not include, for example, correlated
  systematic uncertainties in the energy scale.}
\label{fig:LE}
\end{figure}

The ratio of the background-subtracted \nuebar\ candidate events,
including the subtraction of geo-neutrinos, to no-oscillation
expectation is plotted in Fig.~\ref{fig:LE} as a
function of L$_{0}$/E. The spectrum indicates almost two cycles of the
periodic feature expected from neutrino oscillation.

In conclusion, KamLAND confirms 
neutrino oscillation, providing the most
precise value of \DeltaMSq\ to date and improving the precision of
\ThetaParam\ in combination with solar $\nu$ data. The indication
of an excess of low-energy anti-neutrinos consistent with an
interpretation as geo-neutrinos persists.

% Acknowledgements
The KamLAND experiment is supported by the Japanese Ministry of
Education, Culture, Sports, Science and Technology, and under the
United States Department of Energy Office grant DEFG03-00ER41138 and
other DOE grants to individual institutions. The reactor data are
provided by courtesy of the following electric associations in Japan:
Hokkaido, Tohoku, Tokyo, Hokuriku, Chubu, Kansai, Chugoku, Shikoku and
Kyushu Electric Power Companies, Japan Atomic Power Co. and Japan
Nuclear Cycle Development Institute. The Kamioka Mining and Smelting
Company has provided service for activities in the mine.

\end{document}